\documentclass[a4paper,12pt,reqno,superscriptaddress]{revtex4}

\usepackage[centertags]{amsmath}
\usepackage{amsfonts}
\usepackage{amssymb}
\usepackage{amsthm}
\usepackage{newlfont}
\usepackage{stmaryrd}
\usepackage{mathrsfs}
\usepackage{euscript}
\usepackage{graphicx}
\usepackage{enumerate}
\usepackage{tikz}
\usepackage{pgf}
\usetikzlibrary{positioning,fit,calc}
\usetikzlibrary{arrows,automata}
\usepackage{wrapfig}
\usepackage{subfigure}
\usepackage{amscd}
\usepackage{hyperref}

\theoremstyle{plain}

\theoremstyle{definition}

\theoremstyle{remark}

\newcommand{\bbC}{{\mathbb C}}

\newcommand{\bbI}{{\mathbb I}}

\newcommand{\opunit}{\text{1}\kern-0.22em\text{l}}

\DeclareMathAlphabet{\mathpzc}{OT1}{pzc}{m}{it}

\newcommand{\id}{\textrm{d}}

\begin{document}

\title{{\bf No information or horizon paradoxes for Th.~Smiths \\ }}

\author{Christian Maes, Instituut voor Theoretische Fysica, KU Leuven}

\begin{abstract}
\underline{Th}e \underline{S}tatistical \underline{m}echanician \underline{i}n \underline{th}e \underline{s}treet (our Th.~Smiths) must be surprised upon hearing popular versions of some of today's most discussed paradoxes in astronomy and cosmology.  In fact, rather standard reminders of the meaning of thermal probabilities in statistical mechanics appear to answer the horizon problem (one of the major motivations for inflation theory) and the information paradox (related to black hole physics), at least as they are usually presented.  Still the paradoxes point to interesting gaps in our statistical understanding of (quantum) gravitational effects. \end{abstract}

\maketitle
\baselineskip=20pt

\section{Initial comments}
Paradoxes hit us with some mystery.  The simpler and the shorter their formulation, the more appealing and the more frequently they are discussed on scientific fora and at meetings.  Paradoxes challenge our understanding and often insist on conceptual clarity.  They have appeared about everywhere in scientific discourse, some having a long history.  An ancient example is Democritos' paradox, as described for example in {\it Nature and the Greeks} by Erwin Schr\"odinger \cite{pear}:  take a cone, or make it something more tasty like a pear, and slice it anywhere parallel to its base. The two circular faces thus produced must have the same area; they just perfectly fit together.  But, if they are equal in size, how could the pear ever get its cone-shape?  That paradox touches at the foundations of statistical mechanics because it deals with the relationship between atomism as a physical theory and the continuum nature 
of macroscopic objects.   It witnesses to the long struggle, as apparent also in Zeno's paradox, of reconciling the continuum with discreteness, of describing physical limits in a correct mathematical language when starting from corpuscular concepts of nature.  Another famous now 20th century example is the twin paradox originating in special relativity and first described by Paul Langevin.  The problem with the twin has little or nothing to do with statistical mechanics  (except when one would discuss the influence of traveling on metabolic processes) and no confusion on atomism or on thermal aspects can arise. It just teaches us about the essential structure of Minkowski space-time and the invariant meaning of proper time.   However, when combined with quantum field theory, the problem touches on the one of accelerating travelers observing black-body radiation at the Unruh temperature, having the same form as the Hawking temperature of a black hole.  Time after time then paradoxes induce discussions on the physical interpretation of the formalism of our best theories or on possible extensions or unifications.\\

Sometimes however history repeats itself, and paradoxes are brought forward that have been answered long ago.  It probably means that the paradox is still much alive and deserves further elaboration when formulated in a new context.  It could also  suggest that simple things have been forgotten or that sloppy thinking has gone unnoticed.  Whatever is the case, the problems or paradoxes that are discussed below are very present in today's scientific communication.  They appear in introductions of talks; there are books written about them, and the web has very many entries
repeating naive mistakes against some of the foundations of statistical mechanics.  It is likely that there are different and much more interesting and deeper issues associated to the horizon problem and to the information paradox, but the present contribution discusses these paradoxes as encountered most often in the streets of physics.  And the suspicion of Th.~Smiths is then unavoidable: can there not just be some misunderstanding here of basic statistical mechanics?\\

As the title suggests, the present paper is not truly a specialized scientific one in the more traditional sense.  It is more popularizing and perhaps provocative at times, hoping it can stimulate discussions on fundamental questions in high--energy physics from the point of view of statistical mechanics.  A more technical paper concentrating on sharper discussions of the information paradox is under construction (joint work with Wojciech De Roeck).

\tableofcontents

\section{Horizon problem}
The usual statement of the so called {\it horizon problem} is at best naive and at worst fundamentally misconceived.  Here is the Wikipedia version (20 February 2015) \cite{wik}: {\it The horizon problem is a problem with the standard cosmological model of  the Big Bang which
was identified in the late 1960s, primarily by Charles Misner.
It points out that different regions of the universe 
have not ``contacted'' each other because of the great
distances between them, 
but nevertheless they have the same temperature
and other physical properties.
This should not be possible, given that
the transfer of information (or energy, heat, etc.) can occur,
 at most, at the speed of light.}

\subsection{What is equilibrium?}\label{2nd}
The notion of thermodynamic equilibrium has various aspects, and it obviously depends on considered spatio-temporal scales and types of observations.  Forgetting much of these last subtleties and speaking operationally, equilibrium is the thermodynamic condition of macroscopic systems where there is a homogeneous temperature, chemical potential(s) and pressure. In that way, an equilibrium system has no systematic currents or collective motions of energy or particles, subsystems are again in equilibrium and their dynamical condition is that of detailed balance, reversibility and not showing any difference between evolution forwards or backwards in time.  Many more features can be added, and depending on the physical situation (e.g. on how the system is open to the environment) various ensembles can be used to describe equilibria mathematically, each governed by their own thermodynamic potential (and corresponding variational principle) and associated Gibbs machinery with its thermal probabilities as introduced also by Maxwell and Boltzmann.  The theory and its mathematics are very well developed, a highlight of 20th century physics, including the fundamental understanding of phase transitions, critical phenomena and possible instabilities related to long range interactions (such as the Jeans instability for gravity), \cite{grav,kies}.\\
There is however a deeper statistical understanding of the equilibrium condition, which starts from basic observations on the particularities of systems composed of a very large number of constituents.  To  unveil already the key-ingredient: the law of large numbers is to be expected to play a fundamental role in any description in terms of additive quantities involving a massive number of terms. \\

I start from the simplest set-up for the phase space of classical mechanical systems at fixed energy $E$,
volume $V$ and particle number $N$.  It represents all the allowed microscopic states $X$ (positions and momenta of all the particles in the system) and indeed we take as working hypothesis that all states $X$ on that energy surface are equally probable (microcanonical distribution).  That Liouville measure is rather natural, unbiased as it is and invariant for the mechanical evolution.
Indeed the Hamiltonian evolution defines a flow in that space, which is reversible and incompressible.\\
  We can try to classify the macroscopic conditions of the system by first defining a number of macroscopic quantities and to see what are the possible values.   The transition from microscopic states $X$ to macroscopic condition (values of macroscopic variables) is formalized by a many-to-one map $X\longmapsto M(X)$. Generally $M(X)$ is achieved through spatial averaging (as when computing a density) or by counting averages (like the fraction of particles with a given property). That map roughly induces a partition on the constant energy surface, dividing it into patches of all states $X$ that have the same macroscopic value $M(X)$. The largest patch is the condition referred to as equilibrium.
  \begin{figure}[h]
  \centering
  \includegraphics[scale=0.5]{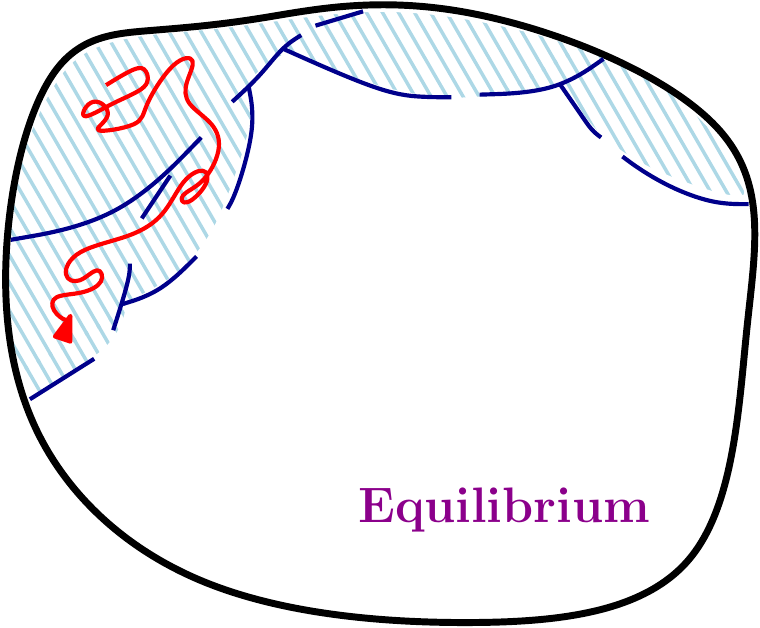}
  \end{figure}
   That induces a partition of the phase space in ``rooms'' or phase space regions that distinguish between macroscopic conditions.\\
 It goes without saying that the coarse graining is physically inspired and the choice of macroscopic variables is not completely arbitrary: we prefer macroscopic descriptions that are sufficiently simple and yet in a way, are dynamically closed (e.g. giving rise to first order dynamical evolutions on hydrodynamic scales).  Here there is  a role for the specific dynamics but in fact already earlier the Hamiltonian has entered as we are fixing the energy and we only consider microscopic states with that energy.\\
      The rest of the story is statistical and is based on another property of the relevant macroscopic quantities: they are arithmetic averages over space or over the various particles of local quantities.
For example, the density in mass or energy can show a macroscopic profile as made from the various local concentrations of masses or energy.  It is here that the law of large numbers starts to play:  Typically, when randomly selecting a phase space point, it belongs to
the phase space region of  ‘thermal equilibrium,’ 
 where macroscopic quantities take their equilibrium values. It is by very far the largest room in the phase space, as visualized in the figure above. For many-particle systems the equilibrium region will be overwhelmingly huge compared to the other (nonequilibrium) regions and in that way the equilibrium values are {\it typical} values just as in the law of large numbers. No details about the system or its dynamics have been specified yet except that we fixed the energy and that we want our macroscopic description to be relevant and (in a sense) complete (for macroscopic autonomy). Irrespective of that, it tells us that equilibrium is the most probable condition from the macroscopic point of view.\\

 Perhaps we can as well remember the 
 \textbf{Boltzmann entropy}, given by $$S(E,V,N;m) = S(m)=k_{B}\log |\{X: M(X)=m\}|$$ where $|A|$ of a phase region $A$ denotes its Liouville volume, always given the constraints on $N$, $E$ and $V$. It quantifies the plausibility of a macroscopic condition, in the sense that
 \[
 \frac{\text{Prob}[m]}{\text{Prob}[m']} = e^{[S(m) - S(m')]/k_B}
 \]
 where $m,m'$ stand for macroscopic conditions, outlook, values,...  Take here into account that the entropy $S$ scales with the number of particles, so that an entropy change of order 1 Joule/Kelvin is easily reached in the kitchen.  Such an increase of entropy easily gives rise to a factor $\exp 10^{20} \simeq \infty$ for the ratio of probabilities.\\
 I repeat that equilibrium (as sensed for example by homogeneous temperatures) is in that sense quite the opposite of ``special.'' 
 Here is how the Master himself was saying that:\\
 {\it One should not forget that the Maxwell distribution is.... 
  in no way a special singular distribution
 which is to be contrasted to infinitely many more 
 non-Maxwellian distributions; rather it is
 characterized by the fact that 
 by far the largest number of possible
 velocity distributions have the characteristic properties of the
 Maxwell distribution,
  and compared to these there are only a relatively
 small number of possible distributions that deviate significantly
 from Maxwell's.}
 (Ludwig Boltzmann, 1896)\\
 Wait, do not read on without having understood that citation.

\subsection{Usual formulation of the horizon problem}

The horizon problem belongs to the original theoretical motivations for inflationary cosmology. The general feeling there is that, as in the standard cosmological model no causal process can establish thermal relaxation within the presently observable universe, we need inflation (a period of accelerated expansion) to account for the {\it observed} thermal homogeneity.  There may well be other reasons to believe in inflation or, what would be even better, solid observational evidence for the process of inflation, but that is not discussed here.\\
Here is the usual presentation of the horizon argument, \cite{mac,nasa}: \underline{claim 1)} there is something special and even very implausible about having equal temperatures  in distant not causally connected regions of our universe, as was measured with relative fluctuations of $10^{-5}$ on the WMAP for cosmic microwave (black body) radiation and confirmed by the Planck ESA mission.  \underline{claim 2)} that specialness of equal temperatures can be removed and hence understood by an inflation scenario (somehow pushing back in time the Big Bang) as that would allow thermalization via causal contact to have taken place after all.  We refer to  the NASA-page \cite{nasa} for a summary of that standard formulation.  In the following section I address the above two claims \cite{oth}.

\subsection{Solution}
I start with a summary of the answer. Equal temperatures are the rule in equilibrium and equilibrium is typical. If indeed temperature is found to be almost  homogeneous, then there is no need and indeed no use in explaining that via thermal relaxation unless you know for sure that the temperatures were not equal at a previous time, if defined at all.   In fact, the universe never was and still is not in equilibrium concerning the gravitational degrees of freedom (including products as ourselves), but there is no reason to doubt its large scale homogeneity in temperature at any moment in its evolution.\\

 Here is my evaluation of the reasoning based on the above Boltzmann picture:  \underline{ad claim 1)} No, on the contrary, there is nothing special about equal temperatures. In fact, equal temperatures are typical for all regions which are solely constrained to conservation of energy. If we imagine the universe with the standard cosmology according to a Friedmann-Lema\^{\i}tre-Robertson-Walker geometry
$\id s^2 = - \id t^2 + a(t)^2 (\id x^2 + \id y^2 + \id z^2)$ with at an initial time $t=t_i$ shortly after the Big Bang an arbitrary matter distribution with a given total energy, then we can and should expect uniform temperature all over. That is just the statement that equilibrium is typical. \underline{ad claim 2)} As a matter of logic, thermalization makes the universe less special so that thermalization cannot explain specialness; the universe would have needed to be more special before.  In other words requiring thermalization is not only not needed; it is worse than useless.\\
The above comments concerning the horizon problem are not original; they have been around for some time, and have been written down more or less in the same way by a number of people; see in particular the analysis by Roger Penrose \cite{pen} on the horizon problem and personally I have greatly benefitted from discussions with Shelly Goldstein; see also \cite{sean}.\\

We can also add that a rough calculation (in communication with Frederik Denef) based entirely on equilibrium fluctuation theory shows that the fluctuations as measured by the WMAP are bigger (not {\it smaller}) than expected.  The calculation uses that for black body radiation the specific heat scales like $C\propto VT^3$.  For volume $V$ we take one pixel of the WMAP CMB image, which is a volume much bigger than a cube light year. For $T$ we can take a temperature of about  1000 Kelvin at the time of CMB emission.  The relative standard deviation in energy scales like $\sigma_E/E \propto 1/\sqrt{C}$.  Hence we get $\sigma_E/E \ll 10^{-30} \ll 10^{-5}$.\\  It should not surprise us however that there is such a deviation as clearly gravitational degrees of freedom are extremely important here. The universe was indeed very special at the Big Bang.  As far as we know and can reasonably assume (also based on the fact that all verified calculations on nuclear synthesis and chemical reactions in the early universe are based on standard equilibrium thermodynamics), there appears to have been a thermal equilibrium but not for the gravitational degrees of freedom 
which prefer a clustered (macroscopic) matter distribution.  Ever since, the universe is relaxing its gravitational degrees of freedom along the Einstein equation of general relativity.  At any rate, if we want to know why the universe was initially so very special for its macroscopic gravitational/geometric condition, we need completely different arguments from those alluded at in the horizon problem. 
If we enquire why it was thermally in equilibrium, there is and remains Boltzmann's answer: that is only normal; it is a matter of counting.\\

Note that the above is not saying that contact or dynamics would not play a role in the \emph{approach} to equilibrium.  Derivations of e.g. diffusive behavior are not at all simple.  But even there, it is not {\it just} the dynamics that matters and statistical reasoning on typicality of initial conditions will play a crucial role.  Maxwell characterized a proposed reduction of the Second Law of thermodynamics 
to a theorem in dynamics, with\\ 
{\it as if any pure dynamical statement would submit
to such an indignity} (Letter to Tait, 1876).  The truth of the second law is as in a statistical theorem, 
{\it of the nature of a strong probability... not an absolute certainty}
like dynamical laws. (Foreword in a book by Tait, 1878).

\subsection{Statistical mechanics of gravity}
In the above I have in Boltzmann's picture not addressed specifically the case of long range interactions (such as in Newtonian gravity).  The ideas are indeed not different, but the consequences look different than for a dilute gas.  As I said, gravity typically leads to clustering and that is not at all in contradiction with Boltzmann's ideas, on the contrary.  Note here that some problems can be created (not solved) by taking the so called canonical ensemble for treating a system of particles with (very) long range interactions; as is well-known the canonical and microcanonical ensemble need not always be equivalent and there are special problems with the physical meaning of the canonical ensemble when the difference between bulk and boundary fades.  The microcanonical treatment for systems with gravity does not present any special conceptual differences with that of dilute gases, but of course equilibrium looks totally different.  We have thus also emphasized that the  hot Big Bang exactly by its supposed matter homogeneity is a very nonequilibrium state of affairs.  Yet, an even better understanding of gravity from a statistical mechanical point would certainly be welcome \cite{wein}.  More specifically I have in mind that a good unison between general relativity and (nonequilibrium) statistical mechanics has not at all been found. The book by Richard Tolman, {\it Relativity, Thermodynamics and Cosmology}, written in 1934 must be revisited \cite{tolm}, at last.  In fact only in the last decade or so a renewed interest has been observed in kinetic relativistic gas theory and  that is still restricted to special relativity.  So the horizon problem should get us moving, and remains valuable, if not as motivation for an inflation scenario, then to get us to start thinking about questions as:\\

\noindent - What are the kinetic constraints in the geometric relaxation of the universe to equilibrium?\\
- What are the roles of expansion and gravitational instability in a statistical mechanical description of  the universe?\\
- What are the relevant and correct macroscopic variables in a geometric theory of gravity? How to quantify here the distance to equilibrium?  Is there an associated Boltzmann entropy which satisfies an H-theorem?\\ What is the quantum statistical mechanical equilibrium of a gravitational system --- Hawking radiation?\\
- How can one formulate the balance equations of irreversible thermodynamics in general relativity?

\section{Information paradox}

The so called information paradox consists of multiple questions and problems related to the construction of a quantum theory of gravity.  It turns out that our understanding today is not optimal, and in particular that shows up in various specific attempts that run into inconsistencies. 
That is very normal for a scientific domain in full development, but it is not very specific.  In fact it is not easy to get a sharp and precise version of the paradox.  That black hole information paradox appears in many different versions, changing in time and depending on the source \cite{para}.  What it is generally accepted to imply is that our usual effective descriptions do not seem to work. It is heard that our effective quantum field theory in which we have good reasons to trust, produces conflicting or inconsistent results.
Surely it could very well be that the quantum field theory that is usually applied there is inappropriate, or must be extended; the claim that follows below is much more conservative:  even \emph{within} the usual scheme of quantum field theoretical understanding of Hawking radiation, the problem is false; there is no inconsistency for Th. Smiths. Or, the usual arguments leading to the paradox are not well founded statistical mechanically.  Does it mean that we do understand a quantum theory of black holes?  No, but let us concentrate on the real problems, and the information paradox if formulated in any sharper way does not appear to be one of them.

\subsection{Entangled pair creation under unitary evolution}\label{form}
Here is again a Wikipedia version (20 February 2015) \cite{wik}:  {\it
Physical information seems to ‘disappear’ in a black hole, 
allowing many physical states to devolve into the same state,
breaking unitarity of quantum evolution.
From the no-hair theorem, one would expect the
Hawking radiation to be completely independent of the 
material entering the black hole. Nevertheless, if the 
material entering the black hole were a pure quantum state, 
the transformation of that state into the mixed state of 
Hawking radiation would destroy information about the 
original quantum state. 
This violates Liouville's theorem and 
presents a physical paradox.}

These previous lines do return in many of the (more) relevant references, but I do not include here a list of the most popular ones.  There is for example the book {\it An Introduction to Black Holes, Information and the String Theory Revolution.  The Holographic Universe}
by Leonard Susskind and Jameson Lindesay \cite{suss}, or the collection of papers in {\it Quantum Aspects of Black Holes}, Ed. Xavier Calmet, Springer Fundamental Theories of Physics, \cite{xa}.  We do not comment on the way entropy or the Second Law are presented there, and move directly to the core of the argument  (basically Chapters 8-9 in \cite{suss} and the paper on the Firewall Phenomenon by  R.B.~Mann in \cite{xa}).  We also do not discuss the precise meaning of the firewall proposal and how its introduction was thought to avoid some ``entanglement paradoxes,'' \cite{fire}.\\

To find a somewhat  precise formulation of the information paradox is not so easy for Th.~Smiths. There are many web-entries, some very instructive as prepared e.g. by Samir Mathur \cite{mat,sam}. A very clear version is in \cite{pol}, also having the advantage of being fairly recent and containing discussions of previous remarks and proposed changes.  My own version of it follows now and I try to summarize the useful discussions I had with Wojciech De Roeck; let me cut the story in pieces:
\begin{enumerate}
	\item
First there is the condition of unitarity which is emphasized. That is not much more than to say we want a description starting from microscopic mechanics. 
Whatever happens in the formation process or evaporation process of a black hole, the evolution is unitary, mapping the initial wave function to intermediate and final wave functions with a so called unitary $S$-matrix.  The unitarity is especially important here because it will lead to the mathematical identity \eqref{mid} essential to the paradox.
\item We make the usual splitting of Hilbert spaces between the singularity and the outer black hole; we are just considering now the process of evaporation and radiation being emitted by the black hole.  So we will consider (for short) the inner (singular) and the outer part of the black hole.  There is the interior of the black hole (behind the horizon, region B) and there is the outside or the exterior (region A).  The total system (A and B together) are quantum mechanically described by a pure state, a wave function $\Psi(t)$.  It is unitarily evolved from another pure state $\Psi(0)$ where we put time zero at the beginning of the evaporation process as described by Hawking radiation.  If we want to consider the outside region A, we can integrate out the degrees of freedom in B.  In quantum mechanics, that means we trace out, and, in contrast to classical states, we do not obtain in general another pure state describing the situation in B, but we obtain a density matrix 
\[
\rho_A(t) = \text{Tr}_B \,|\psi(t)\rangle\langle\psi(t)|
\]
Similarly, there is a density matrix $\rho_B(t)$ for the statistical distribution of the interior of the black hole.
The fact that these are density matrices (and not wave functions) arises because the pairs of photons that are created at the horizon are entangled.   They create in other words an entanglement between regions A and B.  Since the state $\Psi(t)$ was pure, it is a theorem that the entanglement entropy of region A equals that of region B.  Mathematically, that is an equality between von Neumann entropies
\begin{equation}\label{mid}
\text{Tr}_A \,\rho_A(t)\log \rho_A(t) = \text{Tr}_B\, \rho_B(t)\log \rho_B(t)
\end{equation}
at all times $t$. The von Neumann entropy of the density matrix in the outer region must be equal to the von Neumann entropy of the interior black hole. 
\item 
Consider the entanglement entropy of the outer region A (left-hand side in \eqref{mid}).  Can we estimate that?  Here is an important statement --- let us call it the statement of {\it increased entanglement}:  entanglement just increases with further pair creation.  The reason and calculation which is given of that {\it increased entanglement}  is that (a) the Hawking radiation is thermal, and (b) the radiation is additive in pair creation. An observer external to the black hole will see a thermal state characterized by a density matrix and moreover purity of that state is never to be restored even when all the black hole is evaporated.   The computation of increased entanglement is based on that thermal equilibrium distribution for black body radiation (at the Hawking temperature). We are now at page 17 in \cite{pol}.  In the words of the ``Fuzzball person'' \cite{sam}: {\it Thus the entanglement cannot go down ever, and thus the
information cannot emerge in the Hawking radiation.}\\
The increase of the left-hand side of \eqref{mid} is taken from the thermal entropy, which is then increasing alright with every creation.  A detailed calculation can be found on pages 90--93 in the book \cite{xa} in the paper {\it The Firewall Phenomenon} of R.B.~Mann.  To be sure, there is an upper bound on that entanglement entropy, at first growing linear in the steps of pair creation, and then saturating. (In \cite{Bekenstein:1980jp} Bekenstein showed that  there is an upper limit on the amount of entropy (and thus information) one can store in a ``chunk of space-time" of a certain radius; a time-dependent version is found in \cite{Bousso:1999xy}.)  Also, there can easily be imagined corrections to thermality, but one shows that the increase of entanglement is stable; see again p93 in \cite{xa}, or p6-8 in \cite{sam2} based on \cite{sam3}.\\
Secondly and moreover, the calculation for increased entanglement uses that the cumulative pair creation works additively in the entanglement entropy.  Every new pair creation adds an elementary unit to the entanglement entropy. 
\item
Meanwhile, what happens to the region B?   Well, by the Hawking radiation the black hole evaporates and its entropy starts to become smaller and smaller.  The radiation in A grows and region B gets smaller.  At a certain moment (beyond the so called Page time), the region B is getting smaller and smaller and therefore its entanglement entropy  which is always smaller than its real (thermodynamic) entropy must also decrease to zero.  If there is almost nothing left of the black hole, then its entropy gets very small and hence  the right-hand side of \eqref{mid} must decrease to zero.
\item
 Now comes the paradox.   In the equality \eqref{mid}, the previous lines just showed the right-hand side goes to zero. But then the entanglement entropy in region A also decreases in time $t$ which contradicts the {\it increased entanglement},  that the radiation remains entangled, that the density matrix $\rho_A(t)$ never restores purity as was inferred before from taking it thermal.
\end{enumerate}

 \subsection{Thermality does not imply increased entanglement}
 A large number of ``solutions'' and ``answers'' have been proposed; too many to include all of them.  I start by mentioning some of them very briefly; some are rather involved and sometimes get very technical.  Compared with the simple line of arguing above, in general they appear somewhat off--target.\\
 The whole description is of course based on calculations that are approximate.  Hawking's calculation that showed that black holes emit thermal radiation \cite{Hawking:1974sw} uses a semi-classical calculation, say quantum field theory with curved background (in fact, quantum theory of a scalar field in the background of a large classical black hole).  Moreover  the calculation uses locality for example in assuming the decomposition of the Hilbert space structure corresponding to regions A and B.  Nevertheless I would be very surprised that these assumptions and approximations lead to such disaster.  After all, the size of the horizon can be arbitrarily weakly curved when the mass of the black hole is large enough. I do not believe that quantum gravity effects can be relevant at such large length scales.
 \\  What is probably more valid if one wants to criticize these effective theories, is that we had better used \emph{open effective field theory}, i.e., effective field theory of systems that are not closed.  Of course, then one must explain well the ultimate and cogent reasons for using these open system theories, but that is basically a problem of statistical mechanics we are used to and some elements of it will be discussed in Section \ref{thm}.\\
 The suggestion or solution of ``forgetting degrees of freedom'' has been entertained  in e.g. \cite{sam2,sam3}; see also \cite{Mathur:2005zp}, in which Samir Mathur explains why some common beliefs do not resolve the apparent puzzle, which sharpens significantly the nature of the information paradox as originally stated by Hawking.  It appears that the solution of the quantum information paradox is not to be found in these ideas of ``approximate'' thermalization, as I already have mentioned in the point 3 above for {\it increased entanglement}.\\
 A further objection  to the paradox  (again to come back in Section \ref{thm}) states that the description of the black hole space-time by the Schwarzschild metric is truly based on an exchange of limits.  A collapsing shell of matter only becomes Schwarzschild in an infinite time limit. At all finite times it slightly differs.  Still I do not believe that it is the good answer to the firewall phenomenon described above.  The solution is much simpler to state.\\
 
 Let me no longer postpone giving the answer of Th. Smiths: the thermality (approximate or not) of Hawking radiation does \underline{not} imply the {\it increased entanglement} nor does the cumulative nature of pair creation.  Just from rereading the scenario in 5 acts of Section \ref{form}  it is quite clear that the statement of {\it increased entanglement} must be very false --- that is a matter of logic if one accepts all the rest. The entanglement entropy just goes to zero, indeed both right-hand side and left-hand side of \eqref{mid}.  So the arguments that lead to that presumed (entanglement) entropy increase is wrong. Why, how so?\\
 Indeed, even a pure state can very well look like a thermal state for all practical purposes and for local observations; the von Neumann entropy or the Shannon entropy are not continuous; see e.g. \cite{dob}.  It is not because two density matrices very much resemble each other locally, that their von Neumann entropies cannot be drastically different, and the usual argument for {\it increased entanglement} is just and only based on that wrong premise.\\
 The real issue thus concerns the nature of the thermal description.  What is really the meaning and the status of these density matrices and  probabilities that seem to enter our description of being thermal?  How seriously should we take them? The answer is, not too much.  To give a trivial example: the Liouville equation for a mechanical system implies that the Shannon--von Neumann entropy of a density matrix (or phase space distribution) does not change in time, even though it can on the appropriate space--time scale or for some class of observables be considered thermal.  So it then describes equilibrium for many practical purposes and yet its Shannon-von Neumann entropy (like in \eqref{mid}) does not at all equal the thermodynamic entropy (which itself however can be written as the Shannon-von Neumann entropy of a thermal density matrix); see Appendix \ref{sha} for a toy-example. That density matrix (solution of the Liouville-von Neumann equation) is for example also quite irrelevant for a more microscopic understanding of the Second Law.   Similarly, a wavefunction for a many-body system can statistically reproduce a thermal state when evaluated for local observables, but that does not imply that the two are equal in all possible senses.  Certainly, their (global) von Neumann entropy need not be equal. One should realize here that a wavefunction or a density matrix for a quantum many-body system contains much more information than for example the one-particle statistics.  So it is not because you can reproduce locally the black body radiation spectrum for a big density matrix  that it would entail that its von Neumann entropy is even approximately equal to that of the corresponding thermal state. Mathematically, entropy is not a continuous functional (with respect to such weak metric).\\
Moreover, even when not inserting (wrongly) the ``mathematically thermal'' condition (i.e., even when not literally using thermal density matrices), one must still avoid a second mistake: that the pair creation as such which is additive, need not lead to increased entanglement.  The reason is simply that the strict correlations with the inner degrees of freedom in the black hole easily get lost; see Appendix \ref{bag} for a simple toy-example.\\

 The previous answer is just a detection of where was the mistake in the reasoning of Section \ref{form}.
 To Th.~Smiths it would be like saying that the Shannon entropy of the Liouville evolved probability distribution equals the (real) thermodynamic entropy.  That is certainly false, even though there may appear good reasons to say that the distribution is thermal indeed. Hawking radiation is essentially pure but that does not require a breakdown of effective field theory. Nevertheless that answer does not describe the mechanism of purification or how to calculate the true and correct entanglement entropy \eqref{mid}. 
 That is a much more detailed and complicated question, which is typically not even tackled for much simpler systems.  But it also seems to take us in the wrong direction, the quantum-mechanical description of black holes and of singularities probably requires much more interesting and important challenges, and it may well be that the answer depends strongly on
 the version of quantum mechanics that one considers, \cite{war}.

 \subsection{Starting from no-hair, but birds do fly}\label{thm}
 
 Not so very long ago, speaking of entanglement was considered (bad) philosophy \cite{ph}. In those times, the information paradox was formulated quite differently from what is written above, but still with a similar flair as people have in general had less reservations to associate the word information with entropies.\\
  Hawking's theorem in \cite{Hawking:1974sw} showing that black holes emit thermal radiation is then combined with another important fact about classical black holes, namely the ``no-hair" theorem: \textit{A stationary four-dimensional solution of the Einstein-Maxwell equations (in Lorentzian signature) is uniquely characterized by its mass ($M$), angular momentum ($J$), electric charge ($Q$), and magnetic charge ($P$).} See the reviews \cite{Chrusciel:1994sn,Robinson}.\\
 The thermal radiation by black holes together with the ``no-hair" theorem seem to suggest that one can take a pure state with charges $(M,Q,P,J)$, evolve it in time to form a black hole and then observe a thermal radiation coming out of the black hole. That is in a nutshell what was the original formulation of the ``information paradox'' since it seems to suggest that one can evolve a pure state into a thermal one thus breaking unitarity. (The version of the paradox presented above in Section \ref{form} is more specific and more recent.)\\
 Also here (and in relation with the firewall phenomenon) various ``answers'' have been formulated.  There is for example the claim (contained e.g. in \cite{Chowdhury:2007jx}) that using AdS/CFT (a topic that got recently much attention) one can understand exactly how information leaks out of an asymptotically AdS black hole.  Those answers and in particular those based on AdS/CFT dualities may very well be correct, but is there not a much simpler issue here, which is felt intuitively by Th.~Smiths?\\

The reply here must be that the no-hair theorem already supposes a stationary limit-situation.  But in such limiting regimes dissipative effects easily arise and are compatible with pre-limit unitary evolution.   Very often, to make things very sharp, we take some thermodynamic limit after which the (reduced) description is particularly simple.  It can for example suffice to give energy, density and volume to completely describe a gas, or to give temperature and magnetization to describe a magnet etc.  That is similar to the status of these no-hair theorems in black hole physics: they involve limiting procedures and considerations of asymptotic stationary behavior both in time and in degrees of freedom.  In these same limits and for the appropriate variables, autonomous dissipative evolutions appear rigorously as mesoscopic and macroscopic behavior with no other input than Hamiltonian or unitary microscopic laws. The first example of that was probably the proof of the Boltzmann equation for a dilute gas \cite{lan}. Perhaps that is related to the fuzzball proposal for black holes \cite{Mathur:2005zp}. The basic idea is that the black hole  has microstates to account for its thermodynamic entropy and the typical microstate is some ``fuzzy" space--time which has features on the scale of the horizon and at asymptotic infinity looks like Minkowski space. One can then somehow average out over these ``fuzzy" geometries and obtain the black hole as a coarse-grained effective description of the physics.\\
  
Perhaps we are witnessing something similar to what happened with classical mechanics when confronted with hydrodynamics and thermodynamics.  Here are two historical examples --- both concerned with Theorems, mathematical rigorous work that has lead to paradoxes that are however solved by understanding their complete irrelevance for the true situation (which mathematically amounts to exchanges of limits).\\
  The first one dates from 1752 and contains what became known as the d'Alembert paradox \cite{dal}, meaning the rigorous conclusion from classical mechanics that birds cannot fly. (More precisely, d'Alembert saw that both drag and lift are zero in potential flow which is incompressible, inviscid, irrotational and stationary.)  The resolution is of course found in the emergence of viscosity either from Hamiltonian mechanics or from Eulerian hydrodynamics, as explained through the validity of the Navier--Stokes equation (1822--1845), or more generally from microscopic (statistical) derivations of the Second Law (as I had a chance to hint at in Section \ref{2nd}).\\ 
The second one is a paper in 1889 by Henri Poincar\'e where he shows that no monotonically
increasing (entropy) function in time could be defined in terms of the canonical
variables in a theory of $N$-body Hamiltonian dynamics \cite{poin}.  In that way he added to the so called irreversibility paradox, and indeed its solution shows that that Poincar\'e theorem is quite irrelevant; see for example \cite{she}.

\section{Conclusion}
To what comes of the horizon problem and the information paradox to the streets of statistical mechanics, there appears the following reply:\\
For the horizon problem: seeing equilibrium aspects such as homogeneous temperature of the background radiation cannot be a problem if there is no reason to suppose it was different before.
The gravitational degrees of freedom appear very far from equilibrium, yet not contradicting the large scale almost equal temperatures without previous causal contact.\\
For the information paradox:
unitary evolutions become effectively dissipative in a reduced description, but formal paradoxes can easily arise when interchanging limits of time and thermodynamic limit.  Macroscopic steady state descriptions are by their nature approximations, valid in some limiting regimes of spatio-temporal scales.  That a statistical distribution is thermal for local observations or for all practical purposes does not need to imply that its Shannon--von Neumann entropy coincides with that of the thermal distribution.  In particular, the statement that the Hawking radiation ``is'' thermal, does not include that the entanglement entropy as seen by the outside observer is the Bekenstein--Hawking entropy or is not ultimately decreasing to zero.  Moreover, the cumulative or additive pair creation does also not imply a never-decreasing entanglement entropy of the radiation as the inner degrees of freedom of the black hole, while shrinking, can become much more internally correlated. (See some toy-examples in the Appendices.)\\

There have been times of great mutual interactions between researchers in statistical mechanics and in field theory.  Several ideas on dynamically broken symmetries and on collective phenomena have been shared and they shaped the landscapes of condensed matter and elementary particle physics alike. The renormalization group, universality, effective coupling, etc. remain key-concepts in all of theoretical physics. Today there appear new opportunities, also for Th. Smiths, for joined initiatives and efforts at Institutes for Theoretical and Mathematical Physics, including work and discussions on a fluctuation theory for gravitation with special challenges regarding the statistical mechanics of the Big Bang and black holes.\\

\noindent {\bf Acknowledgment}\\
\baselineskip=10pt
{\scriptsize{\bf These ideas have been presented in the physics colloquium at the ENS Lyon on 2 February 2015.  I thank Nikolay Bobev, Thomas Van Riet and especially Wojciech De Roeck at the Institute for Theoretical Physics in Leuven for many discussions and encouragement. I also thank Ward Struyve and Bert Vercnocke for a careful reading and warnings.}}



\appendix
\baselineskip=20pt
\section{Example of  a ``thermal'' distribution with Shannon-von Neumann entropy very much differing from the thermal entropy}
\label{sha}
I repeat here some ingredients of the quantum Kac ring model \cite{ext,prag} to show how a quantum unitary evolution gives rise to ``effective'' dissipative behavior for some thermodynamic observables.  In fact the model has similar semi-classical aspects as found for quantum field discussions in a curved back ground.  Yet, the Kac model is so transparent that all calculations become very simple.\\
Consider a ring $\Lambda$ of a large number $N$ of sites each carrying a quantum spin $\eta(i)\in \bbC^2$ and a classical background variable $g_i\in \{0,1\}$. We characterize that background by the average $r= \frac 1{N}(g_1+\ldots g_N)$. For the quantum dynamics we choose a unitary matrix $V$ on $\bbC^2$ and define $U^N\eta =$
\[
\left(g_NV\eta(N) + (1-g_N)\eta(N),g_1V\eta(1) + (1-g_1)\eta(1),\ldots,g_{N-1}V\eta(N-1)+ (1-g_{N-1})\eta(N-1)\right)
\]
which is extended by linearity to be unitary on the Hilbert space $\bbC^{2N}$ (given the background $g$ which is unchanged in the dynamics).\\
For observables we take one-site matrices, the Pauli matrices $\sigma_\alpha$ in the three directions $\alpha=1,2,3$ and magnetization operator
\[
M_\alpha^N =\frac 1{N} \sum_{i=1}^N \sigma_\alpha(i)
\]
where the $\sigma_\alpha(i)$ are copies of $\sigma_\alpha$ at site $i$.
As initial condition we can essentially take any situation which is concentrating on a particular magnetization vector $m$, by which we mean an initial state $\omega^N$  for which the expectations $\omega^N\left[M^N_\alpha\right] \rightarrow m_\alpha$ for large $N$ (convergence in mean).  In other words, magnetization is initially well-defined. We now evolve unitarily
\[
\omega_t^N \left[\cdot\right]=  \omega^N\left[(U_N)^{-t}\,\cdot\,(U_N)^t\right] \quad \text{ from } \omega^N_0=\omega^N \quad \text{ with } t=0,1,2,\ldots
\]
and we are interested in the large $N$ statistics.  Remember that $\omega^N$ has magnetization $m$. One can then prove \cite{ext,prag} that, as $N\uparrow +\infty$
\begin{equation}\label{mag}
\omega^N_t\left[\,F(M_\alpha^N)\,\right] \rightarrow  F((\phi_tm)_\alpha), \quad \alpha=1,2,3
\end{equation}
for all continuous functions $F$, where $\phi_t$ is a (truly-)dissipative autonomous dynamics in the magnetization-space.
In other words, the unitarily evolved quantum state $\omega_t^N$ can be seen as the result of a dissipative evolution when restricted to the magnetization observables.  For large $N$ and any fixed time $t$ the induced statistics on the magnetization is indistinguishable from that obtained from a relaxational dynamics towards equilibrium.\\
We can still say it differently.  Any magnetization vector $m$ gives rise to a $2\times 2$ density matrix  $\nu = \frac 1{2}\left(\bbI + m\cdot \sigma\right)$. We define the dissipative evolution, with previously introduced background parameter $r$,
\[
\hat{\phi}(\nu) = r V \nu V^\dagger + (1-r)\nu
\]
which after $t$ steps takes $\nu\rightarrow \nu_t = \hat{\phi}^t(\nu)$.  Then $\phi_t(m) = \text{Tr}[\nu_t \,\sigma]$.  Concerning magnetization observations the theorem \eqref{mag} thus states that for all times $\omega_t^N\simeq \bigotimes^N \nu_t$, disordered with a magnetization evolving dissipatively.\\

Obviously, the von Neumann entropy of $\omega^N_t$ is constant in time $t$, but that of $\nu_t$ is not.  Understandably, the Tr$[\rho \log \rho]$ expression for $\omega^N_t$ and for $\bigotimes^N \nu_t$ differ enormously.  The latter is truly thermal for large time $t$ with a temperature that is directly derived from the initial condition, and its Boltzmann entropy, Gibbs entropy, von Neumann entropy and thermodynamic entropy all coincide.   Nothing like that is true for $\omega_t^N$ and yet, it is ``thermal'' as well for all purposes of magnetization.

\section{Example how additive ``correlated pair creation'' does not lead to ever-increasing correlations}
\label{bag}
The point of the following little story is to show that even when we enforce strict thermalization (in a mathematical sense), still the correlations need not keep increasing.  Here we thus leave quantum unitary evolutions, possibly having in mind an effective description on a subclass of observables as in the previous Appendix \ref{sha}, and concentrate on the relation between additive pair creation and building up correlations.\\

Suppose we have a big container with a total of (a very large even number) $N$  black and white balls.  They are moving around quite chaotically in a container with a small hole leaving room for one ball at a time to escape the container.  Let us count time discretely with one step containing two escapes. (The escapes mimic the pair creation.) At moment $n=0,1,2,\ldots$ a total of $2n$ balls went through that hole. It is organized in such a way that at their escape (say at the hole) another chaotic control repaints the balls either black or white with equal probabilities but so that necessarily, within each time step, the two escaping balls get opposite colors. (The pair escaping at time $n$ has perfect correlation.) Finally, for each escaping pair one ball is returned inside the container and the other ball is collected in a bag.  All that is classical mechanics and there will be no talk of entanglement; rather we speak here about the correlation between the average color in the bag and the average color in the container.  Yet, very similar models can be set up also for quantum unitary evolutions where the correlation is truly measured in terms of entanglement (but not here in the present paper). \\

For the purpose of calculation we thus have $\eta_i(n)=\pm 1, i=0,\ldots,N-n$ at time $n$ for the balls in the container. We take for example black to be +1 and and we call -1 white.  We start at time zero with an equal number of black and white balls.  The total color in the container at time $n$ is
\[
M_1(n) = \sum_{i=1}^{N-n}\eta_i(n)
\]
($M_1(0) = 0$.)  The $i-$label has no permanent meaning; it just orders the balls at each moment.
At all times the bag is as disordered as as series of coin tossing; balls that fell in the bag are black or white, independently and with equal probability.   (They are the analogue of thermal radiation.)   Yet the bag's color is correlated with the container's color, surely increasingly so in an initial period.\\
Let $\sigma_j(n)=\pm 1, j=1,\ldots,n$ denote the colors of the balls in the bag, with total color
\[
M_2(n) = \sum_{j=1}^n \sigma_j(n)
\]
(and here the $j-$label can denote the order in which the balls arrived in the bag, but that is of no further importance). At time zero $M_2(0)=0$ say (no balls in the bag).   At time $n=1$ one ball $M_2(1) = \sigma_1(1)=\pm  1$ arrived in the bag, black or white with equal probability, and in the container $M_1(1)$ has three possible values, $M_1(1) = M_1(0), M_1(0)-2, M_1(0)+2$ with probability 1/2, 1/4 and 1/4 respectively.  At any event, $M_1(1) \,M_2(1) = -[\sum_{i=1}^{N-2}\eta_i(1) + \xi]\,\xi$ where $\xi$ is the color of the ball that was returned inside the container, with $\xi=\pm 1$ with equal probability.  At that moment the balls in the container are reshuffled and again two of them are picked out, colored randomly but differently after which one returns to the container and the other is added to the bag, {\it etcetera}. The correlation at time $n$ when averaged over many runs equals  $\langle M_1(n)\,M_2(n)\rangle = - n(N-n)/(N-1)$. 
There is an initial period (till about time $N/2$) of increased (anti-)correlation after which, time-symmetrically, (anti-)correlations decrease.\\   I repeat that the present example is not taking care of unitarity; it just makes the point that (1) one can easily produce maximal disorder (in the bag), and additively in (2) each time, for each pair, impose strict anti-correlation, while still (3) the correlation is non-monotone in time with the source.  I thank Urna Basu for discussions on that toy-model.


\begin{thebibliography}{99}
\bibitem{wik}
The Wikipedia initiative is often a source of good information and is very much appreciated. I quote it here simply because it thus represents the majority of writings on the subject.  It also avoids embarrassing specific authors.
\bibitem{oth}
There are other claims and formulations under the horizon problem, see again \cite{mac}.  A popular emphasis is on the high correlations between causally not connected regions or between temperature and other physical fields.  I have not been able to find a precise formulation of that ``correlation-paradox'' which is not either the same as the ``more or less'' equal temperature problem, or which is a problem at all (given what one means by correlations which of course need not be causal in general; apple trees and pear trees do not cause their blossoming but the time of their flowering is about equal.)
\bibitem{para}
For example, it is not unusual to hear or read that the paradox is about information, with the key-question whether Hawking radiation carries information or not.  Another emphasis regards the paradox as a contradiction between predictions of semi-classical gravity and expectations for a (good) quantum theory of gravity. Or still others would speak about the unwanted pure--to--mixed transition; the Hawking radiation would become mixed/thermal because of paired particles disappearing in the singularity, thus erasing information, etc. etc.  All of these concerns are recognized and more sharply stated in the recent debates on fire walls.  We come to a more precise formulation in Section \ref{form}.
\end{thebibliography}
\end{document}